\documentclass[a4paper,12pt]{article}
\pdfoutput=1
\usepackage[intlimits,centertags]{amsmath}
\usepackage{amssymb,amsfonts}
\usepackage{bbm}
\usepackage{mathrsfs}
\usepackage{cite}
\usepackage[scale={0.78,0.8}]{geometry}
\usepackage{multirow}
\usepackage{relsize}
\usepackage[subrefformat=parens,labelformat=parens]{subfig}
\usepackage{xcolor}
\usepackage{caption}
\usepackage{cancel}
\usepackage{graphics,graphicx}
\usepackage{float}
\numberwithin{equation}{section}

\newcommand{\be}{\begin{equation}}
\newcommand{\ee}{\end{equation}}

\begin{document}

\pagestyle{empty}
\rightline{IFT-UAM/CSIC-17-065}
\vspace{1.2cm}

\vskip 1.5cm

\begin{center}
\LARGE{UV physics and the speed of sound during inflation}
\\[13mm]
  \large{Francisco G. Pedro$^{1}$  \\[6mm]}
\small{
${}^1$Departamento de F\'{\i}sica Te\'orica
and Instituto de F\'{\i}sica Te\'orica UAM/CSIC,\\[-0.3em]
Universidad Aut\'onoma de Madrid,
Cantoblanco, 28049 Madrid, Spain
\\[8mm]}
\end{center}
\vspace{2cm}
\begin{abstract}
We consider inflation as an effective field theory and study the effects of the addition to the Lagrangian of irrelevant operators with higher powers of first derivatives on its dynamics and observables. We find that significant deviations from the two-derivative dynamics are possible within the regime of validity of the effective field theory. Focusing on monomial potentials we show that the main effect of the terms under consideration is to reduce the speed of sound thereby reducing the tensor fraction, while having little impact on the scalar tilt. Crucially, these effects can arise even when the UV cut-off is well above the inflationary Hubble parameter.
\end{abstract}
\newpage

\setcounter{page}{1}
\pagestyle{plain}
\renewcommand{\thefootnote}{\arabic{footnote}}
\setcounter{footnote}{0}

\tableofcontents
\vspace{0.5cm}

\section{Introduction}

Cosmic inflation is quite possibly the highest energy phenomenon we can probe, involving energies that are orders of magnitude beyond those of the most powerful accelerators. At present, the absence of a strict bound on the speed of sound of the scalar perturbations and  lacking a detection of primordial gravitational waves, we are however unable to determine exactly the energy scale of inflation. If the primordial tensor fraction is not exceedingly small, the situation is likely to change in the coming years, getting us a step closer to determining the energy scale of inflation, and opening the door for a deeper understanding of the early Universe.

Inflation's simplest description is in terms of an effective field theory (EFT) with Einstein-Hilbert gravity minimally coupled to scalar(s) field(s). In order to fit current observations it suffices to consider the canonical two-derivative Lagrangian for the scalars, with a suitably flat potential. The characteristic energy scale of the EFT is the Hubble parameter, which measures the expansion rate of the background spacetime. Within this effective description of the early Universe it is desirable to study the effects of the addition of new interactions to the minimal Lagrangian described above, a procedure at the heart of any EFT approach. These additions can either be corrections to the scalar potential or kinetic term, non-minimal couplings to gravity or higher-derivative operators.

In this paper we will focus on the effects of higher-derivative operators on inflation and its observables. By higher-derivative operators we mean both those operators in the Lagrangian with more than one derivative acting on the physical degrees of freedom, e.g. $\square^2 \phi$, as well as those with more than two powers of a single derivative like for instance $(\partial \phi)^4$. Theories with unconstrained higher-derivative operators, whether in classical mechanics or in field theory, are often problematic \cite{Ostrogradsky:1850fid}. They have more degrees of freedom than their two-derivative counterparts, lack a low-energy bound and feature negative-norm states, see e.g. \cite{Simon:1990ic}, \cite{Jaen:1986iz} and \cite{Antoniadis:2007xc}. These undesirable features of higher-derivative theories disappear when analysing then in an EFT context, since the higher-derivative operators can be consistently treated as a small perturbation on the two-derivative Lagrangian, the resulting equations of motion will still be second order in spacetime derivatives and no ghost-like degrees of freedom appear. This is the approach we adopt when dealing with such terms.

In  the framework of single-field inflation, certain higher-derivative operators are known to induce a sub-luminal speed of sound for the scalar perturbations \cite{ArmendarizPicon:1999rj,Garriga:1999vw}, which modifies the consistency relation and leads to a suppression of the tensor fraction. At present, the observational bounds \cite{Ade:2015lrj} on the speed of sound are rather relaxed, allowing for significant deviations from unity (see also \cite{Zavala:2014bda}). We will exploit this fact in the present paper as we try to probe the effects of higher-derivative operators on inflation. \\

\vspace{0.5cm}
We start by illustrating, in a simple example, how higher-derivative interactions may arise in the context of effective field theories in Sec. \ref{motivation}. Section \ref{sec:analysis} constitutes the bulk of this paper and in it we analyse the effects of various higher-derivative terms on the inflationary observables, working in a regime where these can be treated as a small perturbation of a two-derivative theory. In it we find sizeable deviations from two-derivative dynamics that can affect the inflationary observables in a significant manner. Section \ref{sec:summary} is dedicated to summarising and discussing our findings.

\section{Higher-derivatives from the UV}\label{motivation}

Actions containing higher-derivative terms appear naturally in the context of effective field theories. The only condition for such terms to be generated is the existence of a coupling between the light and heavy degrees of freedom. Let us illustrate how this may happen by considering the following toy model with interacting light ($\phi$) and heavy ($\chi$) scalars: \footnote{We work in units of $M_P^{-2}= 8 \pi G$ and take the metric signature to be ''mostly plus''.}
\be
\mathcal{L}/\sqrt{g}=-\frac{1}{2} (\partial \phi)^2-\frac{1}{2} (\partial \chi)^2-\frac{m^2}{2}  \phi^2-\frac{M^2}{2} \chi^2-\lambda\  \phi^2 \chi \ ,
\label{eq:toy}
\ee
where $m\ll M$.

If one is interested in physics on energy scales $E\ll M$ it is sufficient (and more practical) to study the system in terms of  the low-energy effective action that can be derived from Eq. \eqref{eq:toy}. This effective action can be found by consistently integrating out the degrees of freedom whose mass scale is above the energy scale of interest, in this case $\chi$. To illustrate how the higher-derivative terms appear at low energies it suffices to find the tree level effective action that follows from \eqref{eq:toy}. This can be done by computing light particle scattering amplitudes or by solving the classical equation of motion for $\chi$ and substituting the solution back into Eq. \eqref{eq:toy}. The equation of motion for $\chi$ reads
\be
\left( \square-M^2\right)\  \chi=  \lambda \  \phi^2 \ ,
\label{eq:KHh}
\ee
where $\square \chi=\frac{1}{\sqrt{g}}\partial_\mu(\sqrt{g} \partial^\mu \chi)$.  This equation can be formally inverted to 
\be
\chi=  \lambda \left( \square-M^2\right)^{-1}  \phi^2 \ ,
\label{eq:h}
\ee
where by $\left( \square-M^2\right)^{-1}$ we mean the differential operator that is the inverse of $\left(\ \square-M^2\right)$. Expanding in inverse powers of $M$ one finds
\be
\chi= -\frac{\lambda}{M^2} \left(1+\frac{\square}{M^2}+\frac{\square^2}{M^4}+\mathcal{O}(M^{-6})\right)\phi^2 \ ,
\ee
which allows us to rewrite Eq. \eqref{eq:toy} as
\be
\mathcal{L}_{eff}/\sqrt{g}=\mathcal{L}/\sqrt{g} \big |_{\chi=0} + \frac{\lambda^2}{2 M^2} \phi^4-\frac{2 \lambda^2}{M^4}\phi^2 (\partial \phi)^2 + \frac{2 \lambda^2}{M^6}\left\{(\partial \phi)^4+2 \phi (\partial \phi)^2 \square \phi + \phi^2 (\square \phi)^2 \right\} +\mathcal{O}(M^{-8})\ ,
\label{eq:Leff}
\ee
where partial integration has been used to bring the Lagrangian to this form. We therefore see that the low-energy effective action contains not only $M$ suppressed corrections to the potential and kinetic term of the light field but also the higher-derivative terms we are interested in. While the study of corrections to the effective scalar potential and kinetic term have received considerable attention in the context of inflationary EFTs (both stringy and otherwise), see e.g. \cite{Dong:2010in,Kaloper:2002uj,Burgess:2003zw,Cicoli:2008gp,Buchmuller:2015oma, Bielleman:2016olv}, the higher-derivative corrections, with the exception of \cite{Shiu:2002kg}, have not been studied as widely.  Though generically all types of corrections can be present at low energies, in this paper we will focus solely on the higher-derivative ones with the aim of understanding their effect on the dynamics and observables of inflation. One should however keep in mind that for any given UV realisation, the combined effects of all corrections must be studied simultaneously.

Equation \eqref{eq:Leff} is a specific case of the most general set of four-derivative operators involving only the scalar field, which can be written as \cite{Weinberg:2008hq}
\be
\Delta \mathcal{L}/\sqrt{g}= \frac{f(\phi/\Lambda)}{\Lambda^4}\  (\partial \phi)^4+\frac{g(\phi/\Lambda)}{\Lambda^3}\  (\partial \phi)^2 \square \phi + \frac{h(\phi/\Lambda)}{\Lambda^2}\ ( \square \phi )^2\ ,
\label{eq:DL0}
\ee
where $\Lambda$ is the UV cut-off of the effective theory and $f, g, h$ are dimensionless functions of $\phi/\Lambda$. Whenever $\Delta \mathcal{L}$ is treated as a perturbation on top of the ordinary two-derivative Lagrangian, it can be shown that by using the leading order equation of motion, $\square \phi= V_\phi$, the $g$ and $h$ terms  can be equivalently written as  sub-leading corrections to the kinetic term and to the potential of the scalar field $\phi$ respectively. \footnote{Note that using the leading order equation of motion to eliminate $\square \phi$ terms is equivalent to performing a field redefinition \cite{GrosseKnetter:1993td} involving its derivatives: $\phi\rightarrow\tilde{\phi}(\phi, \partial\phi)$. The order in  $M$ to which the redefinition must be performed depends on the order to which one is truncating the effective action \cite{Gong:2014rna}.} 
 We stress that though the two actions thus obtained are different, they are equivalent, since they give rise to the same physical observables (see the discussions in e.g. \cite{Burgess:2007pt,Gong:2014rna,Weinberg:2008hq}). This leaves
\be
\Delta \mathcal{L}/\sqrt{g}= \frac{f(\phi/\Lambda)}{\Lambda^4}\  (\partial \phi)^4
\label{eq:DL}
\ee
as the unique irreducible four-derivative operator. \footnote{If besides the terms in \eqref{eq:DL0} one also allows for four-derivative terms involving powers of spacetime curvature, the number of operators increases from the three considered above to ten \cite{Weinberg:2008hq,Elizalde:1994sn}. The same simplification procedure applies (where one can now also use the leading order Einstein's equations), and leads to only two further operators at the four-derivative level involving contractions of the Weyl tensor. In this work we will ignore these terms and focus exclusively on corrections of the type of Eq. \eqref{eq:DL}; we point the interested reader to the detailed discussion in  \cite{Weinberg:2008hq}.} The remainder of this paper will be devoted to evaluating the effects of \eqref{eq:DL}, and variations thereof, on inflation.

\section{Inflationary observables and higher-derivative terms}\label{sec:analysis}

The study of higher-derivative Lagrangians in the context of inflationary models is a research program that has been pursued over the last two decades, most notably within frameworks like K-flation \cite{ArmendarizPicon:1999rj,Garriga:1999vw} and DBI-inflation \cite{Alishahiha:2004eh} (see \cite{Stein:2016jja} for a recent analysis). Both constitute significant departures from the standard two-derivative slow-roll and can therefore lead to $c_s \ll 1$. Our approach in this paper is rather more conservative as we will be looking at small perturbations around a two-derivative slow-roll Lagrangian. We take 
\be
\mathcal{L}=-\frac{1}{2} (\partial \phi)^2-\sum_{n\ge2} \frac{f_n}{\Lambda^{4 n-4}} (\partial \phi)^{2n} -V(\phi)
\label{eq:Lstart}
\ee
as our starting point, denoting by $f_n\equiv f_n(\phi/\Lambda)$  the unknown dimensionless functions that may appear in the derivative expansion and by $\Lambda$  the scale of the heavy physics we integrated out in order to get to the effective action  \eqref{eq:Lstart}. In the example of the previous section we can identify $\Lambda=M$. Note that since we are agnostic regarding the exact form of the derivative expansion, we will limit ourselves to a regime where the series converges rapidly such that we can restrict ourselves to considering only the  leading order correction on top of the two-derivative Lagrangian. For concreteness we will take 
\be
f_n \frac{(\partial\phi)^2}{\Lambda^4}<1/10\ ,
\label{eq:bound}
\ee
as the convergence criterion for the derivative expansion.

\subsection{Four-derivative correction}\label{sec:4D}

In this section we assume that the next-to-leading order term is four-derivative and therefore start by analysing the following Lagrangian
\be
\mathcal{L}=-\frac{1}{2} (\partial \phi)^2+ \frac{f}{\Lambda^{4}} (\partial \phi)^4 -V(\phi)\ ,
\label{eq:L4}
\ee
where $f\equiv f(\phi/\Lambda)$. 
As stated above, this regime is a minimal departure from the two-derivative slow-roll, but as we'll see, can have interesting observable consequences in some inflationary models.

This type of model can be easily translated into the formalism of \cite{ArmendarizPicon:1999rj,Garriga:1999vw} whose main results we now review. We start by assuming a flat FRW universe with line element
\be
ds^2=-dt^2+a(t)^2 \left(dr^2+r^2 d\Omega_2\right)\ ,
\ee
where $a(t)$ is the scale factor, whose Hubble parameter we denote by $H=\dot{a}/a$. Upon coupling Eq. \eqref{eq:L4} to Einstein-Hilbert gravity one finds the following equations of motion
\begin{align}
H^2&=\frac{1}{3}\left[\frac{\dot{\phi}^2}{2}\left(1+6 f \frac{ \dot{\phi}^2}{ \Lambda^4}\right)+V(\phi)\right]\ ,
\label{eq:eom1}\\
\dot{H}&=-\frac{\dot{\phi}^2}{2}\left(1+4 f \frac{ \dot{\phi}^2}{ \Lambda^4}\right)\ ,
\label{eq:eom2}\\
\ddot{\phi}&\left(1+12 f \frac{\dot{\phi}^2}{\Lambda^4}\right)+3H\dot{\phi}\left(1+4f \frac{\dot{\phi}^2}{\Lambda^4}\right)+\frac{3 f_\phi}{\Lambda^4}\dot{\phi}^4+V_\phi=0\ .
\label{eq:eom3}
\end{align}

In the analysis of this system it is convenient to define the Hubble slow-variation parameters
\be
\epsilon_H=-\frac{\dot{H}}{H^2}\qquad \text{and} \qquad \eta_H=\frac{\dot{\epsilon}_H}{ \epsilon_H H}\ ,
\ee
which as we'll show can be related to the potential slow-roll parameters
\be
\epsilon_V= \frac{1}{2}\left(\frac{V_\phi}{V}\right)^2\qquad\text{and}\qquad \eta_V=\frac{V_{\phi\phi}}{V}\ .
\label{eq:SRV}
\ee

The effect of the four-derivative term is to induce a non-trivial sound speed for the scalar perturbations given in  this case by
\be
c_s^2=\frac{1+4f\dot{\phi}^2/\Lambda^4}{1+12f \dot{\phi}^2/\Lambda^4} \ ,
\label{eq:cs2}
\ee
whose variation is parametrised by the slow-variation parameter
\be
s=\frac{\dot{c}_s}{c_s H}\ .
\ee
Note that the coefficient of the four-derivative term must be positive, otherwise the speed of sound, Eq. \eqref{eq:cs2},  would be larger than the speed of light.

At the level of linear perturbation theory, the main effect of the higher-derivative terms is to modify the dispersion relation for the scalar perturbations. This in turn implies a modification of the   amplitude of the scalar power spectrum  to
\be
\mathcal{P}_S=\left.\frac{1}{ 8 \pi^2}\frac{H^2}{ \epsilon_H c_s}\right |_{c_s k =a H}\ .
\label{eq:Ps}
\ee
Note that due to the non-standard sound speed, the amplitude is evaluated at the sound horizon crossing: $c_s k =a H$. The scalar spectral index is 
\be
n_s-1=\frac{d \ln \mathcal{P}_S}{d \ln k}=-2 \epsilon_H-\eta_H-s\ .
\label{eq:ns}
\ee
Unlike the scalar perturbations' quadratic action, the tensor's is unchanged by the presence of higher-derivative terms. It then follows that  the tensor power spectrum has the usual amplitude
\be
\mathcal{P}_T=\left.\frac{2}{\pi^2 }H^2\right|_{k=a H}
\label{eq:Pt}
\ee
computed at horizon crossing $k=aH$. In principle scalar and tensor spectra are to be evaluated at different times for models with $c_s\neq1$, since scalar and tensor modes of the same comoving momentum leave the horizon and freeze at different times. In practice the difference only arises at higher order in the slow-roll expansion (see e.g. \cite{Ade:2015lrj} and references therein) and we are free to evaluate the power spectra at $k=aH$ as usual \cite{Garriga:1999vw}. The tensor-to-scalar ratio can then be defined from Eqs. \eqref{eq:Ps} and \eqref{eq:Pt} and takes the form
\be
r=16 \epsilon_H c_s\ .
\label{eq:r}
\ee
In the decoupling limit $\Lambda\rightarrow \infty$ one recovers the standard result of two-derivative Lagrangians: scalar perturbations propagate at the speed of light, $c_s\rightarrow 1$, and the inflationary observables take their familiar two-derivative forms.

A phenomenological and model independent analysis of the effects of $c_s$ on the inflationary observables in the light of the 2013  {\it Planck} data can be found in \cite{Zavala:2014bda}.  As stated above in this work we are interested in minimal deviations from two-derivative slow-roll dynamics, therefore we focus on a regime for which Eqs. \eqref{eq:eom1}-\eqref{eq:eom3} are well approximated by
\begin{align}
&H^2=\frac{V}{3 M_P^2}\ ,
\label{eq:eom1SR}\\
&\dot{H}=-\frac{\dot{\phi}^2}{2}\left(1+4f \frac{ \dot{\phi}^2}{ \Lambda^4}\right)\ ,
\label{eq:eom2SR}\\
&3H\dot{\phi}\left(1+4f \frac{\dot{\phi}^2}{\Lambda^4}\right)+\frac{3 f_\phi}{\Lambda^4}\dot{\phi}^4+V_\phi=0\ .
\label{eq:eom3SR}
\end{align}
Note in particular that the inflaton moves at terminal velocity ($\ddot{\phi}=0$) and that the Hubble parameter is sourced predominantly by the scalar potential, the two main features of slow-roll dynamics.

\subsubsection{Case I: $f(\phi/\Lambda)=1$}\label{case1}

\begin{figure}[h!]
\begin{centering}
\includegraphics[width=0.5\textwidth]{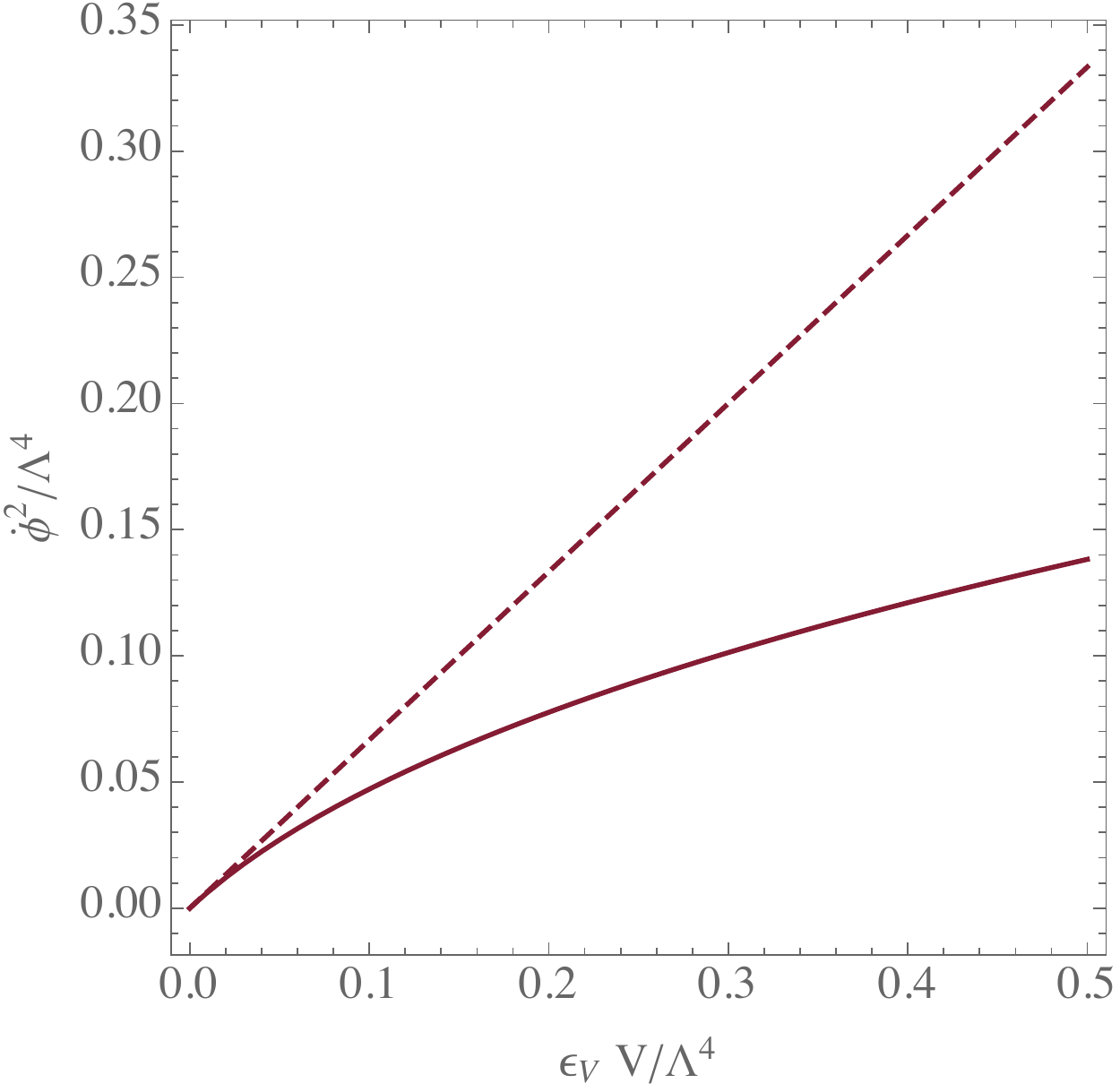}
\caption{Terminal velocity $\dot{\phi}^2/\Lambda^4$: exact result (full line) and leading order approximation (dashed line).}
\label{fig:Delta}
\end{centering}
\end{figure}

We start by analysing the simplest setup $f(\phi/\Lambda)=1$, that follows from having
\be
\mathcal{L}\supset \frac{(\partial \phi)^4}{\Lambda^4}\ .
\label{eq:Lhd}
\ee
As mentioned above, at the background level, the system described by these slow-roll equations is still evolving at terminal velocity ($\ddot{\phi}=0$) with the scalar potential driving the expansion of spacetime. The major difference between the higher-devivative and the two-derivative cases arises from Eq. \eqref{eq:eom3SR}, which determines the terminal velocity in terms of the slope of the scalar potential and of the UV scale $\Lambda$: $\dot{\phi}$ is now determined by a cubic equation \cite{Shiu:2002kg}, whereas in two-derivative slow-roll it was set by the linear relation $\dot{\phi}=-\frac{V_\phi}{3H}$. The solution to the cubic equation can be written as
\be
\frac{\dot{\phi}^2}{\Lambda^4}=\left[\frac{-1+(-3\sqrt{2\ \Delta}+\sqrt{1+18\ \Delta})^{2/3}}{2\sqrt{3}(-3\sqrt{2 \ \Delta}+\sqrt{1+18\ \Delta})^{1/3}}\right]^2\ ,
\label{eq:control}
\ee
where we defined $\Delta\equiv \epsilon_V\frac{V}{\Lambda^4}$. Expanding Eq. \eqref{eq:control} to leading order in $\Delta$ we recover the usual two-derivative slow-roll result
\be
\frac{\dot{\phi}^2}{\Lambda^4} \approx\frac{2}{3}\Delta\: \Leftrightarrow \: \dot{\phi} \approx-\frac{2}{3} \epsilon_V V\ .
\label{eq:SR}
\ee
Figure \ref{fig:Delta} depicts the full result and the leading order approximation for $\dot{\phi}^2/\Lambda^4$.  One sees that the approximation of Eq. \eqref{eq:SR}  only holds for very small values of  $\epsilon_V V/\Lambda^4 \lesssim 0.1$, that is when the cut-off scale $\Lambda$ is sufficiently large relative to the inflationary energy density. This is due to the fact that the solutions to a cubic and a linear equations only match if the coefficient of the cubic term is parametrically small or equivalently due to  the  expansion of Eq. \eqref{eq:control} in $\Delta$ being an alternating series with slow convergence properties. From Fig. \ref{fig:Delta} one also learns that the effect of the four-derivative term is to reduce the terminal velocity of the inflaton, effectively enhancing the friction.

Keeping in mind the above observations, it is still instructive to study the behaviour of the system in an expansion around small $\epsilon_V V/\Lambda^4$ as this will allow us to determine the effect of the four-derivative term on the background evolution and on the inflationary observables. In the higher-derivative slow-roll regime one can show that to leading order in $\epsilon_V V/\Lambda^4$ 
\begin{align}
\epsilon_H&\approx\epsilon_V-\frac{8}{3} \epsilon_{V}^2 \frac{V}{\Lambda^4}\ ,\\
\eta_H&\approx4\epsilon_V-2\eta_V-16 \epsilon_V(\epsilon_V+2\eta_V) \frac{V}{\Lambda^4}\ ,
\end{align}
where $\epsilon_V$ and $\eta_V$ are defined in Eq. \eqref{eq:SRV}. Once again, if one takes the limit $\Lambda\rightarrow \infty$ one recovers the standard leading order relations $\epsilon_H=\epsilon_V$ and $\eta_H=4\epsilon_V-2\eta_V$. The sound speed can also be written in a power series in $\epsilon_V V/\Lambda^4$ to leading order as 
\be
c_s^2\approx1-\frac{16}{3}\epsilon_V \frac{V}{\Lambda^4}\ ,
\ee
while its variation takes the form
\be
s\approx\frac{16}{3}\left(-\epsilon_V+ \eta_V\right) \epsilon_V \frac{V}{\Lambda^4}\ .
\ee
The inflationary observables admit the following expansion 
\be
n_s-1\approx-6\epsilon_V+2\eta_V+\frac{16}{3}  (5 \epsilon_V-3\eta_V) \epsilon_V \frac{V}{\Lambda^4}\ ,
\label{eq:ns}
\ee
and
\be
r\approx16 \epsilon_V\left(1-\frac{16}{3} \epsilon_V \frac{V}{\Lambda^4}\right)\ .
\label{eq:r}
\ee
These are to be evaluated at horizon crossing, $N_e$ efoldings before the end of inflation
\be
N_e=\int_{\phi_{N_e}}^{\phi_{end}} \frac{H}{\dot{\phi}}d \phi \ ,
\ee
when the field takes the value $\phi_{N_e}$. Since the higher-derivative interaction causes a reduction in the terminal velocity we expect, everything else being equal, $\phi_{N_e}$ to be smaller than in the two-derivative case.

From these expressions, in particular \eqref{eq:r}, one sees that if the four-derivative terms are to have any effect on the observables, then the cut-off scale cannot be parametrically larger than the inflationary scale defined as $M_{inf}\equiv V^{1/4}$. By virtue of Eq. \eqref{eq:eom1SR} this can happen while still keeping $\Lambda\gg H$, the condition which allows us consistently integrate out the degrees of freedom associated with the UV scale $\Lambda$.  Note that even if we push the new physics scale $\Lambda$ down towards $M_{inf}$, the derivative expansion is under control at the time of horizon crossing, since $\epsilon_V\sim n_s-1\lesssim
 \mathcal{O}(10^{-2})$.
One may wonder what is the scale $\Lambda$, and how it compares to the remaining scales in the model. Two possibilities come to mind
\begin{itemize}
\item $\Lambda\sim M$, where $M\ (\gg H)$ denotes the mass of the heavy fields one integrated out to get to the effective action of  Eq. \eqref{eq:L4}. In this case we see that $\Lambda/H\sim M/H\gg1$. Temporarily reinstating the factors of $M_P$, this implies that $V/\Lambda^4\sim H^2 M_P^2 /\Lambda^4 \sim (H/M)^2 (M_P/M)^2$.  Note that the first factor is $\gg1$ while the second should be $< 1$ or possibly $\ll1$. It seems that $V/\Lambda^4\sim\mathcal{O}(1)$ is a reasonable possibility in this case.
\item $\Lambda\sim \sqrt{M_P M}$, in this case $V/\Lambda^4\sim H^2 M_P^2/(M_P^2 M^2)=(H/M)^2\ll1$, implying that there is tension between having sizeable deviations from the two-derivative slow-roll observables and at the same time maintaining the derivative expansion under control. Note that this does not mean that interesting deviations from standard case are not possible, but simply that one would need to have an exact, rather than perturbative, description of the kinetic Lagrangian like e.g. in models derived from the DBI action \cite{Bielleman:2016olv,Bielleman:2016grv,Bielleman:2015lka,Ibanez:2014swa}.
\end{itemize}

Another observation is in order regarding the expansions of Eqs. \eqref{eq:ns} and \eqref{eq:r}: since deviations from the two-derivative observables are proportional to $\epsilon_V$, the steeper the potential at horizon crossing, the more pronounced the effects from the four-derivative term will be. This singles out chaotic monomials (as opposed by exponentially flat plateaus) as the ideal arena in which to test these effects. Let us then focus on potentials of the type
\be
V=\lambda_n \phi^n, \qquad\text{with} \qquad n=\{3,2,4/3,1,2/3\}
\ee
and perform a numerical study of the effective action \eqref{eq:L4}. 

\begin{figure}[h!]
	\centering
	\begin{minipage}[b]{0.49\linewidth}
	\centering
	\includegraphics[width=0.92\textwidth]{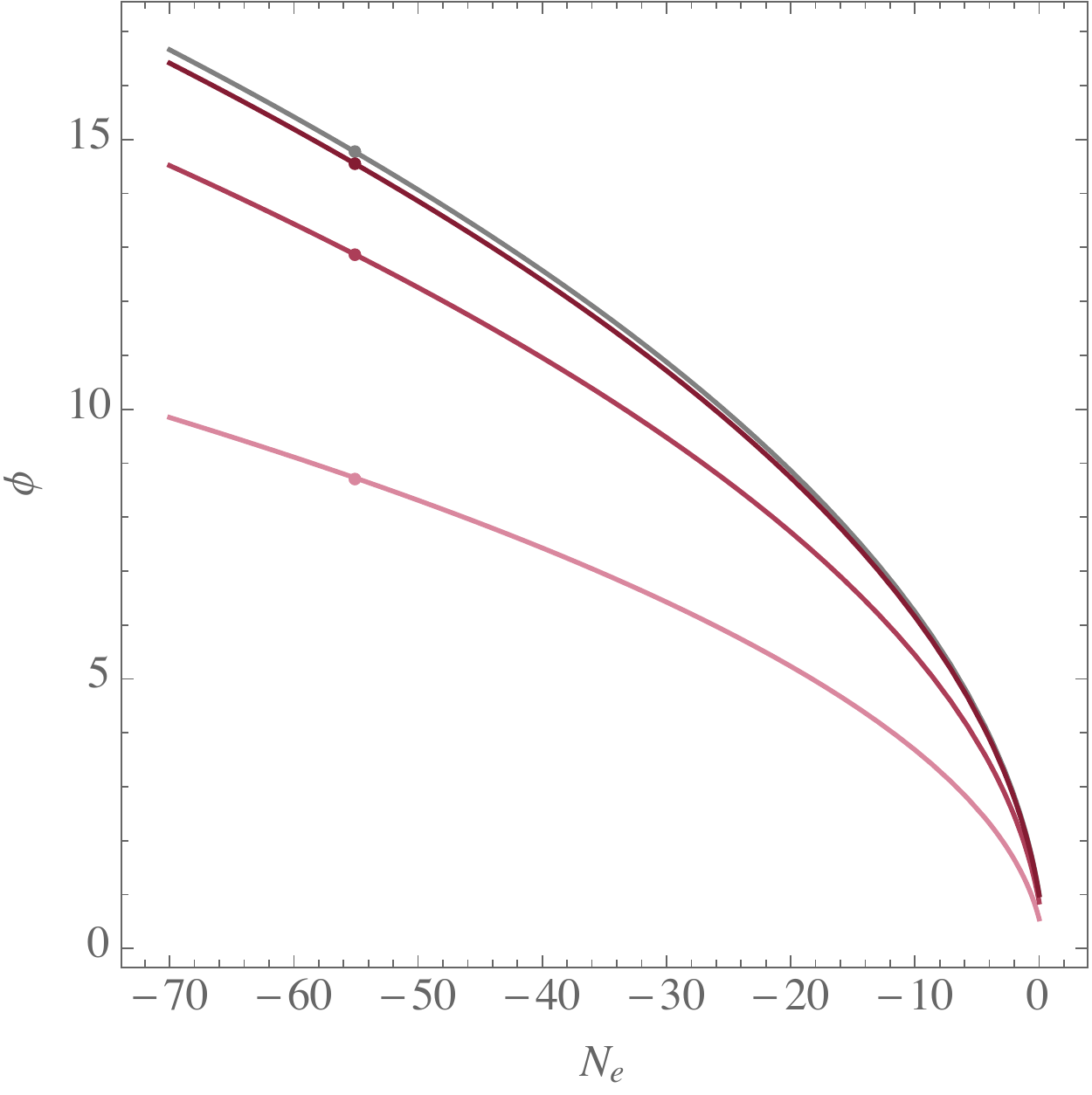}
    \end{minipage}
	\hspace{0.05cm}
	\begin{minipage}[b]{0.49\linewidth}
	\centering
	\includegraphics[width=0.95\textwidth]{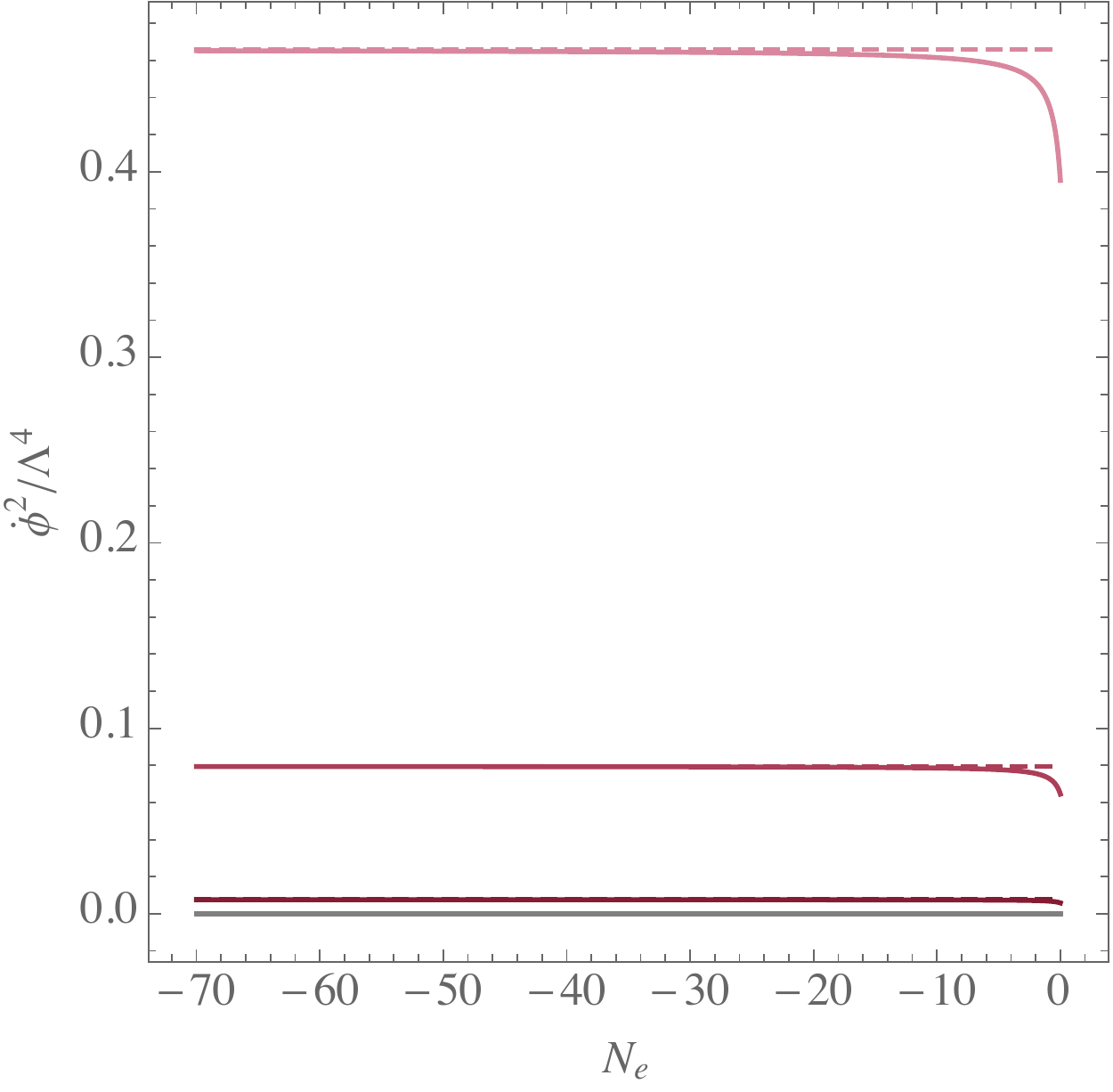}
	\end{minipage}
	\hspace{0.05cm}
\centering
	\begin{minipage}[b]{0.49\linewidth}
	\centering
	\includegraphics[width=0.97\textwidth]{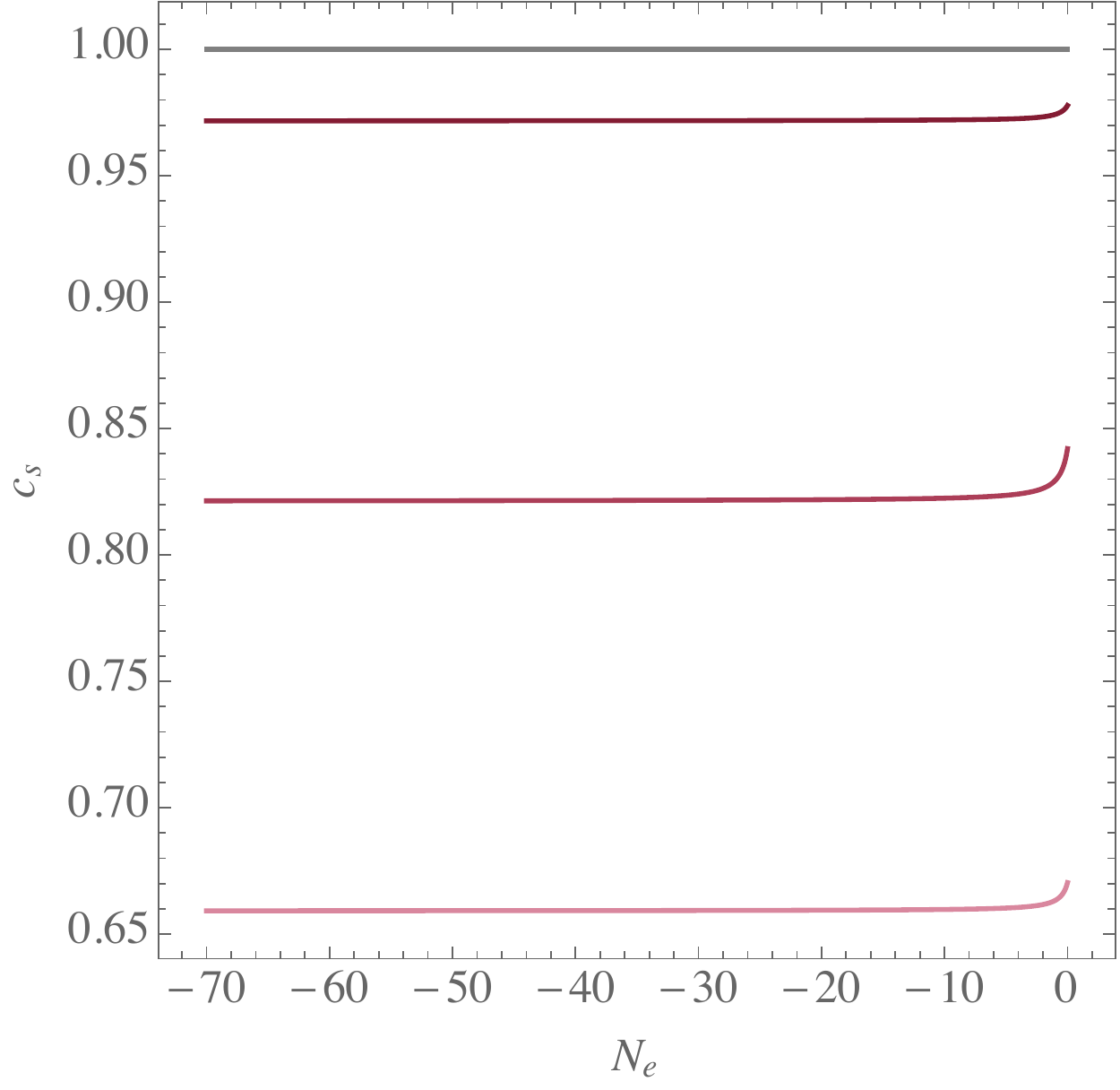}
    \end{minipage}
	\hspace{0.05cm}
	\begin{minipage}[b]{0.49\linewidth}
	\centering
	\includegraphics[width=0.97\textwidth]{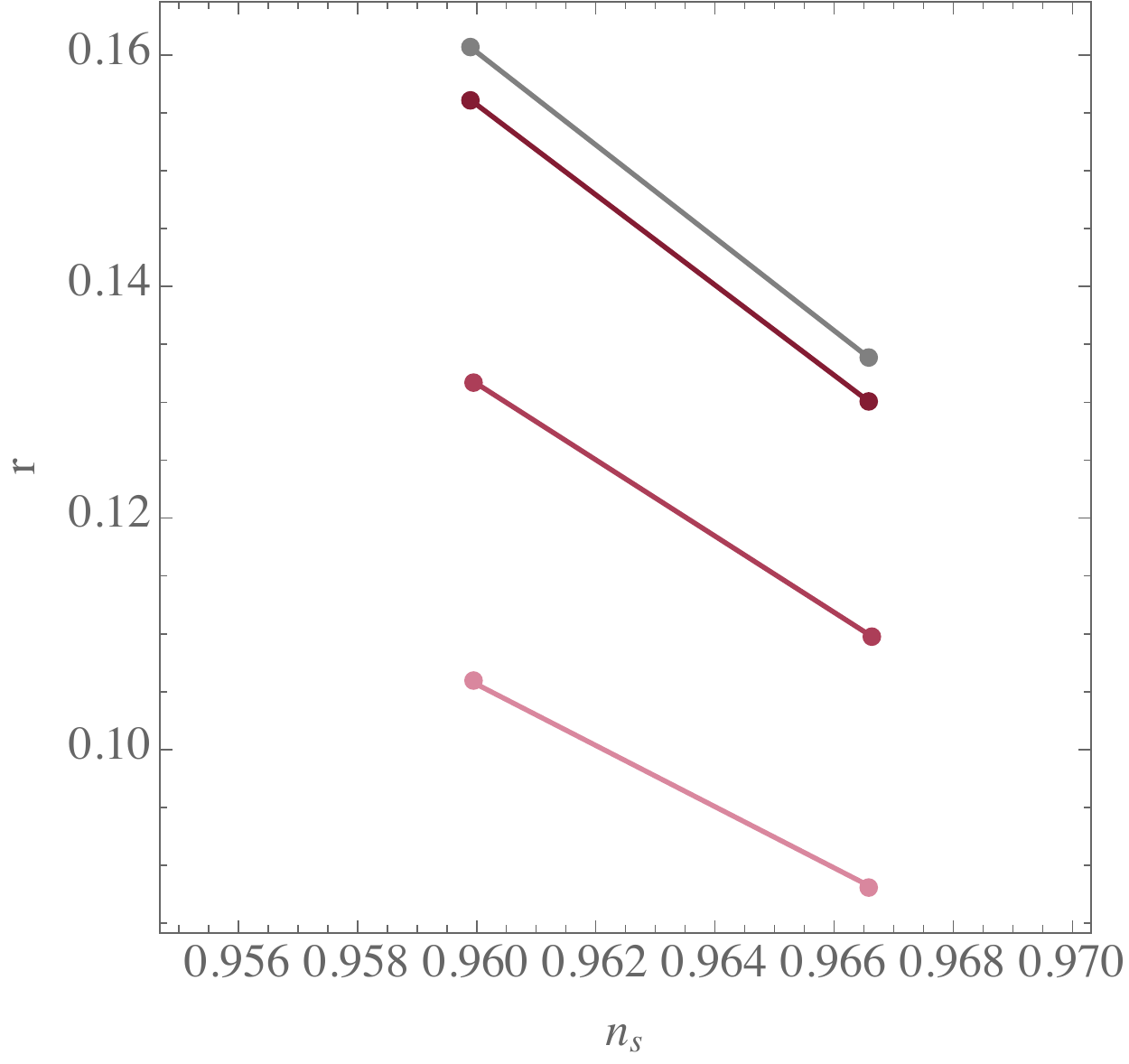}
	\end{minipage}
	\hspace{0.05cm}
\caption{Numerical results for $\phi^2$ inflation for $\Lambda/H\sim\{60,100,200,\infty \}$. Grey represents the standard two-derivative case ($\Lambda=\infty$), darker shades of magenta correspond to larger values of $\Lambda/H$. {\it Upper left panel}: field evolution, {\it Upper right panel} : control parameter of the derivative expansion; solid lines correspond to numerical results, whereas dashed lines are the solution of Eq. \eqref{eq:control}. {\it  Lower left panel}: speed of sound,  {\it Lower right panel} : observable signature in the $n_s - r$ plane for $N_e\in [50,60]$.}
	\label{fig:quadratic}
\end{figure}

\begin{figure}[h!]
\begin{centering}
\includegraphics[width=0.65\textwidth]{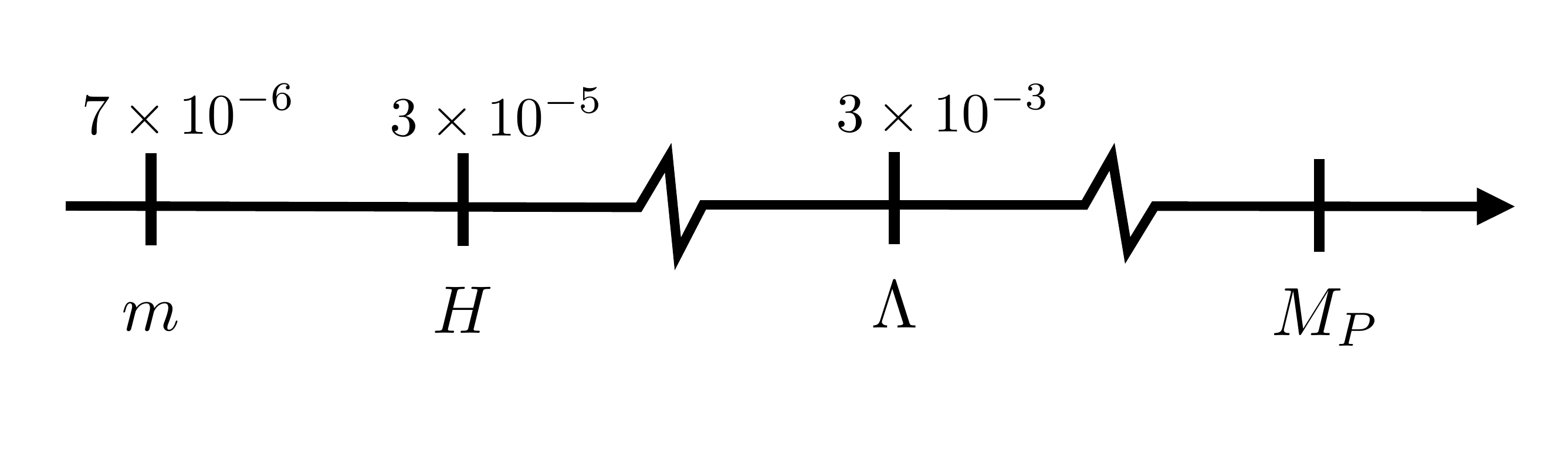}
\caption{Mass hierarchies for $\phi^2$ inflation with $\Lambda/H=100$ in units of $M_p$.}
\label{fig:massScales}
\end{centering}
\end{figure}

In Fig. \ref{fig:quadratic} we present the results for $\phi^2$ inflation. Just like in two-derivative slow-roll, the numerical results are well approximated by the analytical slow-roll limit estimates, with the only difference being that the relation between the velocity and the gradient of the scalar potential is now given by Eq. \eqref{eq:control} rather than by Eq. \eqref{eq:SR}. We scan the cut-off scale $\Lambda$ such that $\Lambda/H\sim\{60,100,200\}$ while ensuring correct normalisation of the scalar spectrum at $N_e=55$.  As expected we see that the smaller $\Lambda$ the more important the effects of the four-derivative term become and the harder it is to keep the derivative expansion under control. From the upper right panel of Fig. \ref{fig:quadratic} we conclude that $\Lambda/H\gtrsim100$ as any lower value of the cut-off will lead to a slow convergence of the derivative expansion and to potentially sizeable deviations from the results presented here. Figure \ref{fig:massScales} shows the mass spectrum for $\Lambda \sim 100 H$ in units of $M_P$.

\begin{figure}[h!]
\begin{centering}
\includegraphics[width=0.6\textwidth]{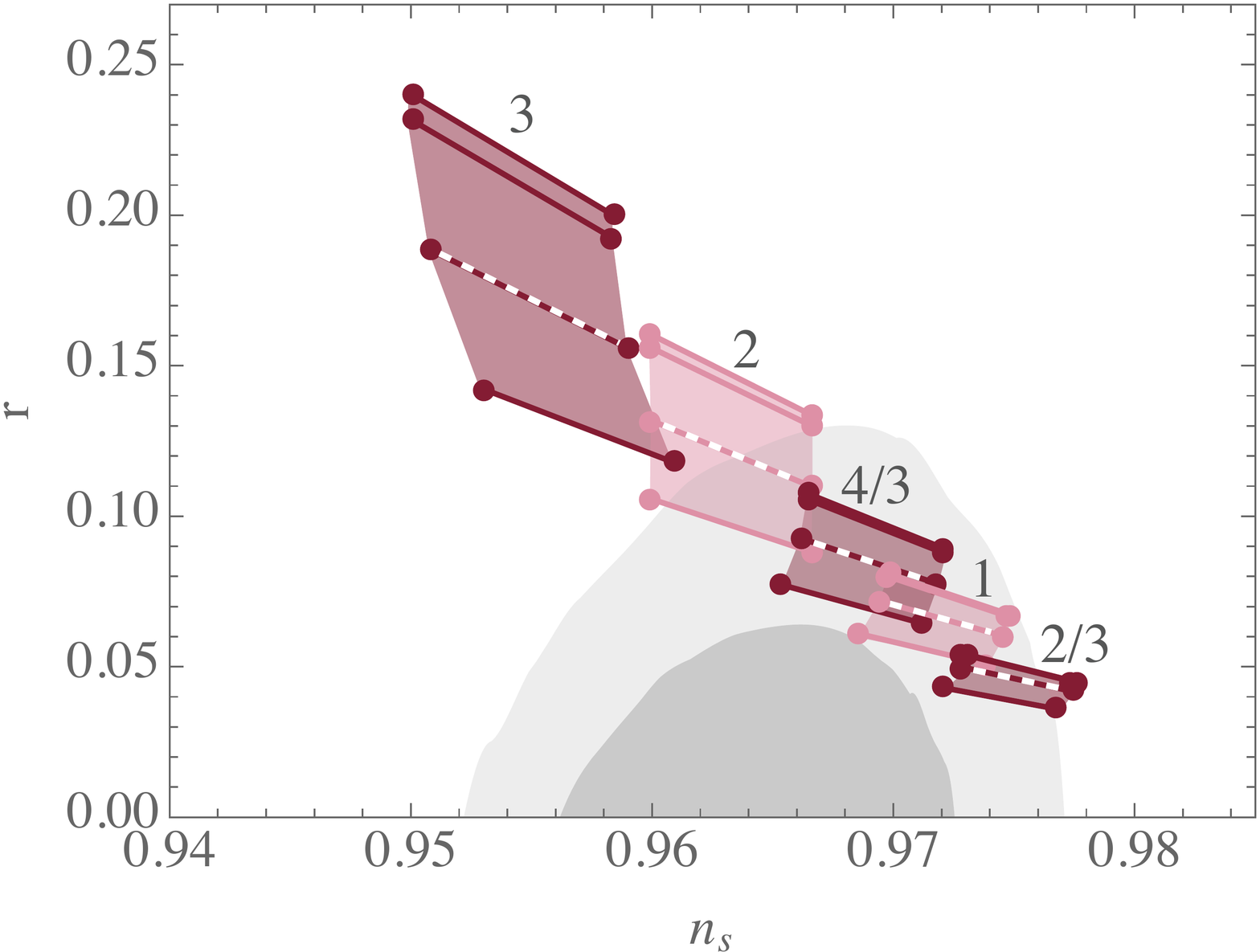}
\caption{Observables of chaotic monomials ($n=3,2,4/3,1,2/3$)  for $\Lambda/H\sim\{\infty,200,100,60\}$. The lower the ratio $\Lambda/H$ the more severe is the effect on the observables. Dashed lines correspond to $\Lambda\sim 100 H$, below which control over the derivative expansion is questionable. Grey shaded regions denote the  {\it Planck}2015 (TT,TE,EE+lowP) constraints \cite{Ade:2015lrj}.}
\label{fig:nsr}
\end{centering}
\end{figure}

We have repeated the analysis for several monomial potentials and present the results in the $\{n_s,r\}$ plane in Fig. \ref{fig:nsr}, again for $\Lambda/H\sim\{\infty, 200, 100,60\}$. The effects of the higher-derivative terms on the observables are qualitatively similar to those of $\phi^2$ inflation: there is a small shift of $n_s$ and a sizeable reduction of $r$. In agreement with the intuition based on the series expansion around large $\Lambda$, we see that the shifts in $r$ are more pronounced for steeper potentials. Finally let us stress that convergence of the derivative expansion throughout the last 60 efoldings translates into the bound  $\Lambda/H\gtrsim100$, like in the quadratic case.

As we have illustrated in Fig. \ref{fig:Delta}, the derivative expansion control parameter $\dot{\phi}^2/\Lambda^4$ is a monotonically growing function of $\epsilon_V V/\Lambda^4$ which exhibits different field dependence depending on the monomial potential one is considering:
\be
\Delta=\epsilon_V \frac{V}{\Lambda^4}= \frac{\lambda_p\  p^2}{2} \phi^{p-2}\ .
\ee
We therefore note that: 
\begin{itemize}
\item for $p>2$,  $\Delta$ decreases as inflation proceeds, so provided the derivative expansion is under control when the pivot scale is exiting the horizon, it will be under control afterwards too,
\item for $p=2$,  $\Delta$ is constant throughout inflation, so control over the derivative expansion around $N_e\sim 60$ ensures control throughout the evolution,
\item for $p<2$,  $\Delta$ increases as inflation proceeds, implying that one needs to ensure control at the end of inflation.
\end{itemize}
This behaviour is the underlying reason why the observables from shallower potentials at horizon exit  are less affected by the higher-derivative terms than those of steeper potentials.

\subsubsection{Case II: $f(\phi/\Lambda)=(\phi/\Lambda)^p$}\label{case2}
\begin{figure}[h!]
\begin{centering}
\includegraphics[width=0.7\textwidth]{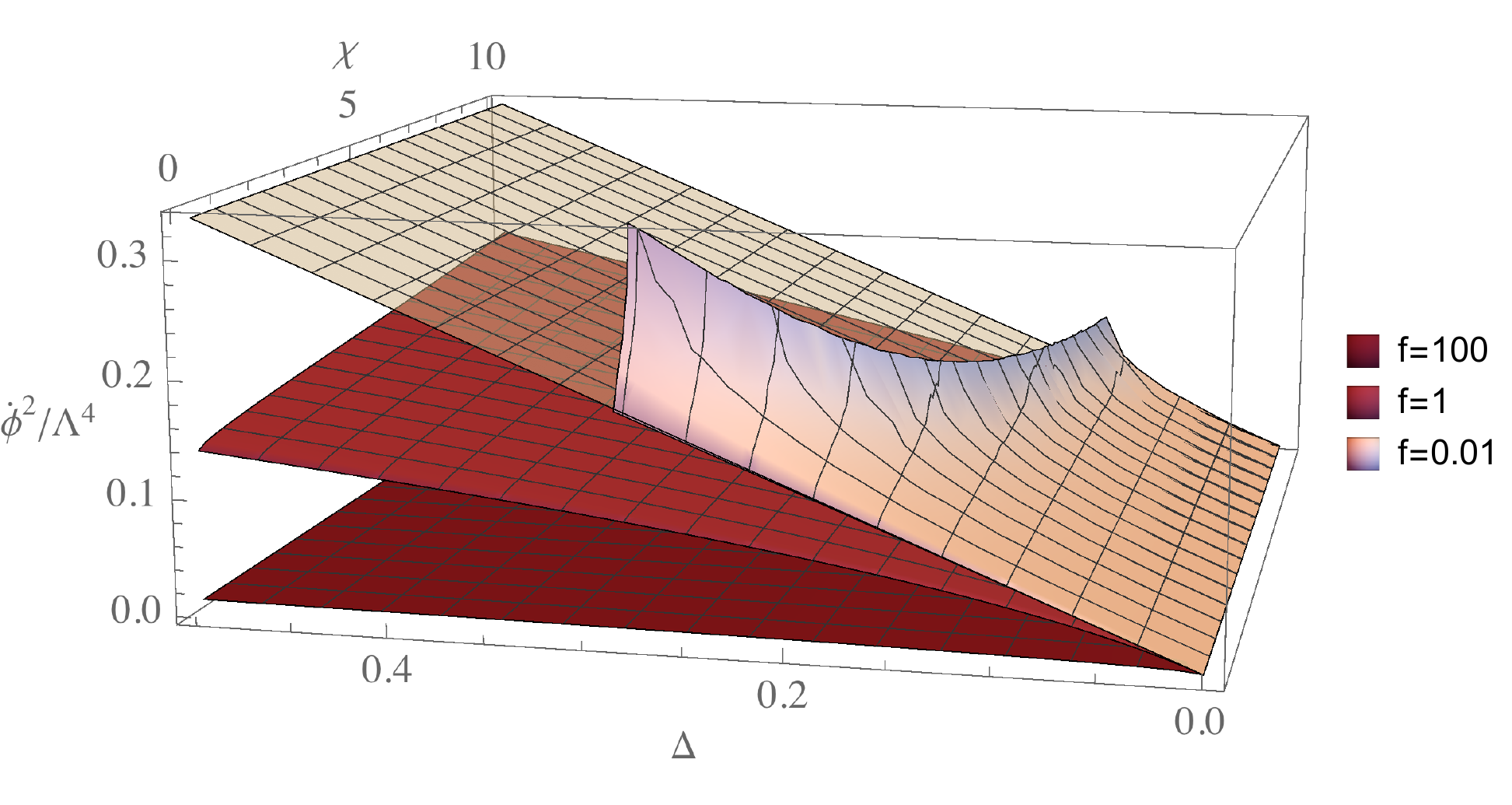}
\caption{$\dot{\phi}^2/\Lambda^4$ as a function of $\{\chi,\Delta,f\}$. The transparent plane corresponds to the terminal velocity of the two-derivative theory, Eq. \eqref{eq:SR}}
\label{fig:Delta_f}
\end{centering}
\end{figure}

We now turn our attention to models in which the leading  higher-derivative correction takes the form
\be
\mathcal{L}\supset \frac{(\phi/\Lambda)^p}{\Lambda^4}(\partial\phi)^4 \ .
\label{eq:Lhdf}
\ee
These correspond to setting $f(\phi/\Lambda)=(\phi/\Lambda)^p$ in the relations of Sec. \ref{sec:4D}. The most obvious modification with respect to the $f(\phi/\Lambda)=1$ case analysed above is that the terminal velocity is now determined by a quartic equation in $\dot{\phi}$ and that there is an extra parameter, $p$. While analytic solutions to this equation can still be found, they can no longer be written in terms of the UV scale $\Lambda$, the scalar potential and its derivatives alone. We therefore choose to proceed numerically, showing in Fig.  \ref{fig:Delta_f}   the solutions of Eq. \eqref{eq:eom3SR} in terms of the dimensionless parameters $\Delta \equiv \epsilon_V V/\Lambda^4$ and $\chi\equiv f_\phi^2 \Lambda^4/(V M_P^2)$ for various values of $f$. Since we are focusing on monomial potentials where $\phi\sim\mathcal{O}(M_P)$ the range of $f$ is determined by $p$: $p>0 \Rightarrow f \gg 1$ while $p<0 \Rightarrow f \ll 1$.

The generic feature of Fig. \ref{fig:Delta_f} is that the terminal velocity that follows from Eq. \eqref{eq:eom3SR} is smaller than in two-derivative slow-roll for most of the  $\{\chi,\Delta,f\}$ parameter space. There is a small region where the inflaton could naively roll faster, but we were unable to find consistent dynamical solutions in that regime. In this respect adding \eqref{eq:Lhd}  or \eqref{eq:Lhdf} to the effective field theory leads to the same effect at the level of the background evolution: the inflaton will roll slower than if these higher-derivative operators were not present. 

In order to get a feel for the effect of the choice of $p$ in Eq. \eqref{eq:Lhdf} may have on the mass spectrum of the system we take a given monomial potential and determine the ratio $\Lambda/H$ for a given value of $c_s$. We find 
\be
\frac{\Lambda}{H}\approx e^{5+ 2 p -p^2/3}\qquad \text{for} \qquad c_s\approx0.9
\ee
at $55$ efoldings is a good approximation for all monomial potentials considered. We get to this result by performing a fit to the numerical results for the various monomials. The only difference between the potentials is the minimum value of $p$ for which the desired sound speed can be achieved without loosing control over the derivative expansion, i.e without violating Eq. \eqref{eq:bound}. As an example for a quadratic potential one can obtain $c_s\approx 0.9$ with $p\ge -1$, lower $p$ values result in a loss of control over the expansion towards the end of inflation, while for $\phi^{2/3}$ the lower bound is instead $p\ge 0$.

\begin{figure}[h!]
\begin{centering}
\includegraphics[width=0.6\textwidth]{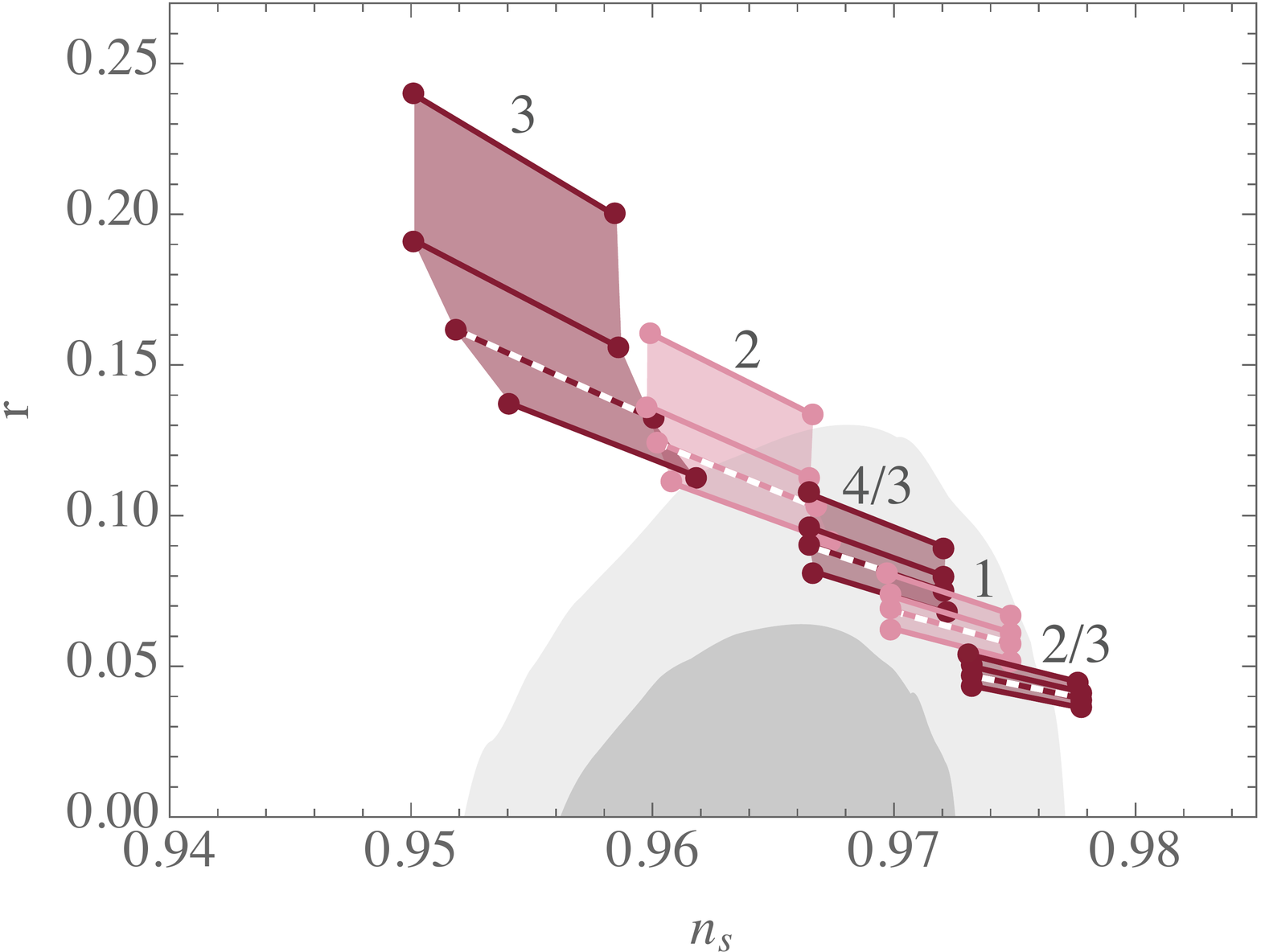}
\caption{Observables of chaotic monomials ($n=3,2,4/3,1,2/3$) for $p=1$ and  $\Lambda/H\sim\{\infty,600,500,400\}$. The lower the ratio $\Lambda/H$ the more severe is the effect on the observables. Dashed lines correspond to $\Lambda\sim 500 H$, below which control over the derivative expansion is hard to achieve. Grey shaded regions denote the  {\it Planck}2015 (TT,TE,EE+lowP) constraints \cite{Ade:2015lrj}.}
\label{fig:nsr2}
\end{centering}
\end{figure}

The effects of \eqref{eq:Lhdf} at the level of the curvature perturbations are qualitatively similar to those reported in the previous Section, with a small shift in $n_s$ and a sizeable reduction in $r$ driven by the non-trivial sound speed as can be seen in Fig. \ref{fig:nsr2} for $p=1$.

\subsection{Higher powers in the derivative expansion}

In the previous sections we have analysed in detail the effect of a four-derivative term in the slow-roll dynamics. For completeness we now look at even higher-derivative interactions of the form
\be
\mathcal{L}_{kin}=-\frac{1}{2} (\partial \phi)^2- \frac{f(\phi/\Lambda)}{\Lambda^{2 n-4}} (\partial \phi)^n\ ,
\label{eq:L_n}
\ee
with $n>2$. Such terms can in principle be the leading correction to the kinetic Lagrangian whenever the UV structure of the theory leads to a fortuitous cancellation/suppression of the  lower order terms.

As in the preceding analysis we do not assume this to be the exact form of $\mathcal{L}_{kin}$, but rather the leading order approximation of an expansion in inverse powers of $\Lambda$. In principle, when one integrates out the heavy fields to find the low-energy effective action a tower of higher-derivative terms will appear. Here we merely analyse the dominant term of such series, which should suffice provided we restrict ourselves to a regime where the series converges rapidly enough, that is provided $f (\partial\phi)^2/\Lambda^4$ is sufficiently small.

\begin{figure}[h!]
	\centering
	\begin{minipage}[b]{0.48\linewidth}
	\centering
	\includegraphics[width=\textwidth]{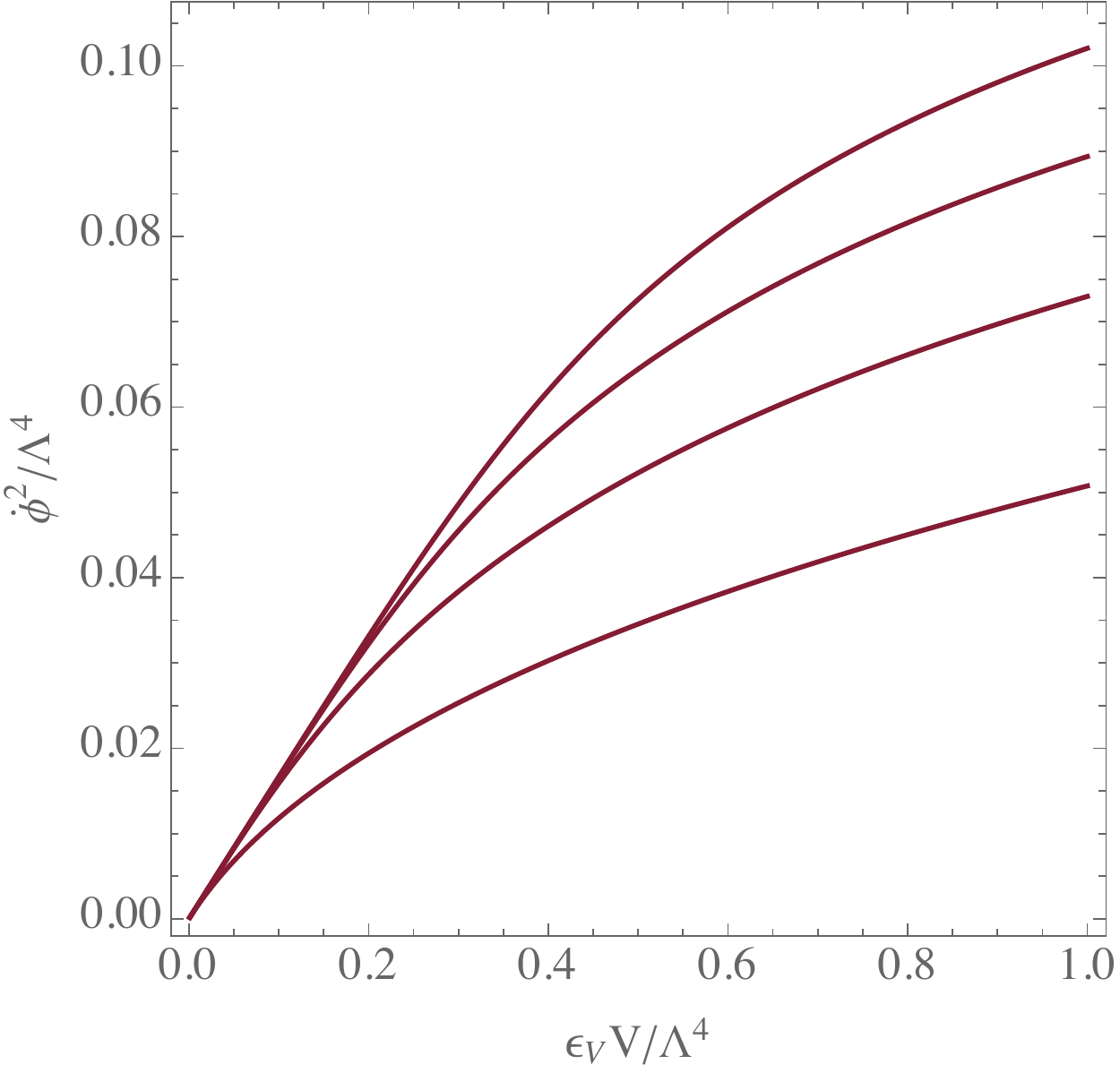}
    \end{minipage}
	\hspace{0.05cm}
	\begin{minipage}[b]{0.48\linewidth}
	\centering
	\includegraphics[width=\textwidth]{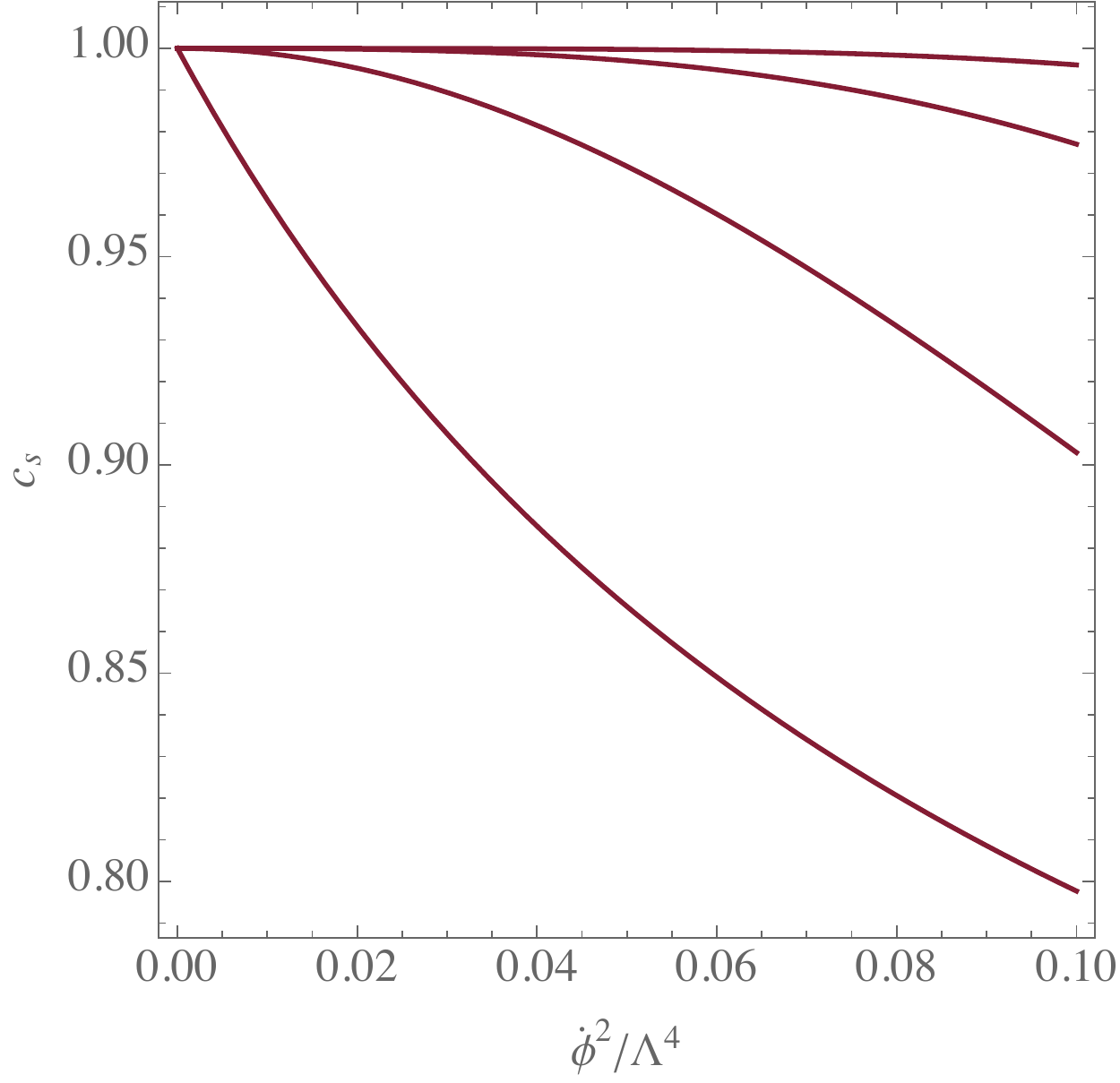}
	\end{minipage}
	\hspace{0.05cm}
	\caption{{\it Left}:  Derivative expansion's control parameter as a function of $\epsilon_V V/\Lambda^4$ for (top to bottom) $n=2,3,4,5$. {\it Right}: Sound speed as a function of the derivative expansion's control parameter for (bottom to top) $n=2,3,4,5$. }
	\label{fig:control_all}
\end{figure}

In light of the results of Secs. \ref{case1} and \ref{case2} we will focus on the $f=1$ case. We can then show that in the slow-roll limit ($\ddot{\phi}=0$), the Klein-Gordon equation for the inflaton reduces to
\be
3 H \dot{\phi}\left(1+n \frac{\dot{\phi}^{2(n-1)}}{\Lambda^{4(n-1)}}\right)=-V_\phi\ .
\label{eq:KG_n}
\ee
Using the fact that the total energy density is well approximated by the potential energy one can further simplify Eq. \eqref{eq:KG_n} to 
\be
\sqrt{\frac{\dot{\phi}^2}{\Lambda^4}}\left[1+n \left(\frac{\dot{\phi}^{2}}{\Lambda^{4}}\right)^{n-1}\right]=-\sqrt{\frac{2}{3}\Delta}\ ,
\ee
where, as before, $\Delta\equiv\epsilon_V V/\Lambda^4$. This allows us to determine how the derivative expansion control parameter depends on $\Delta$ for general $n$. Just like in the four-derivative  ($n=2$)  case of the previous section $\dot{\phi}^2/\Lambda^4$ is proportional to $\Delta$, as illustrated in Fig. \ref{fig:control_all}.

From Eq. \eqref{eq:L_n} it follows that the sound speed is given by
\be
c_s^2=\frac{1+n \ \dot{\phi}^{2(n-1)}/\Lambda^{4(n-1)}}{1+n (2n-1)\ \dot{\phi}^{2(n-1)}/\Lambda^{4(n-1)}} \ .
\ee
By requiring that $\dot{\phi}^2/\Lambda^4<0.1$ we can read off the maximum possible reduction in the sound speed for each value of $n$, finding, unsurprisingly,  that the larger the $n$ the smaller the reduction in $c_s$. These results are illustrated in  Fig. \ref{fig:control_all}.

\subsection{Constraints on the sound speed}

It is well know that at the level of the two point function, there is a large degeneracy between the slow-roll parameters and the sound speed. This can be seen from the expressions for the inflationary observables presented in Sec. \ref{sec:4D}. Such degeneracy makes it hard to derive bounds on $c_s$ from the power spectra alone.  

The higher-order kinetic interactions considered in this work will give rise to interactions in the scalar perturbations' action beyond quadratic level, which can in principle lead to primordial non-Gaussianities for the scalar perturbations, see e.g. \cite{Chen:2006nt, Senatore:2009gt}.  The magnitude of the non-Gaussianities is usually given in terms of the non-linearity parameter $f_{NL}$ which is bounded by CMB observations. By combining the power spectra data and the bounds on $f_{NL}$, the  {\it Planck} collaboration \cite{Ade:2015lrj,Ade:2015ava} is able to put the following bound on the sound speed 
\be
c_s>0.024 \qquad (95\ \% \ \text{CL}) \ .
\ee
This lower bound can be easily  accommodated in the models considered in this paper, given that from the effective field theory point of view one is looking for small UV induced departures from the canonical $c_s=1$. Furthermore attaining such low values of $c_s$ would require working in a regime in which the derivative expansion manifestly does not converge. In such case an exact knowledge of the full derivative expansion would be required, like in e.g. in DBI inflation \cite{Alishahiha:2004eh}.

\section{Summary}\label{sec:summary}

In this paper we have studied the effects of a particular type of higher-derivative operators on the dynamics and observable signatures of cosmic inflation. We have a adopted a conservative approach by working in a regime where these operators are small perturbations on a standard two-derivative Lagrangian. At the level of the background evolution, the principal effect is a decrease of the terminal velocity of the inflaton field that follows from the fact that the Klein-Gordon equation becomes of fourth order in the field velocity. The presence of four-derivative terms is known to induce a sub-luminal sound speed for the scalar perturbations which modifies the amplitude of the scalar power spectrum and consequently the tensor-to-scalar ratio. We find these effects to be more dramatic for steeper inflationary potentials and therefore focused our efforts on the analysis of chaotic monomials. For these models we have shown that the departure from two-derivative dynamics can arise in a regime where the derivative expansion converges rapidly ($f(\phi) \dot{\phi}^2/\Lambda^4\ll1$) and the effective field theory is under control ($\Lambda \gg H$). This can be achieved while keeping the sound speed well above the lower bound set by  {\it Planck}'s analysis of the scalar two and three point functions. In this regime we have found that the inflationary observables are modified: there are small shifts of the scalar spectral index but above all there is a sizeable reduction of the tensor-to-scalar ratio due to a $c_s$ driven enhancement of the scalar amplitude.

In light of these results it would be interesting to pursue this approach further in order to see to what extent the dynamics and the observables of the higher-derivative theory can differ from those of the standard two-derivative case. This could be done by going beyond the single four-derivative term through the inclusion of higher order terms in the expansion and, if possible, by relaxing the strict convergence criterium used here. It would also be instructive to find UV models whose low-energy behaviour is modified by higher-derivative terms, within string compactifications (building on the works of \cite{Ciupke:2015msa} and \cite{Bielleman:2016olv,Ibanez:2014swa}) or otherwise.  We plan to return to these issues in the future.

\section*{Acknowledgments}
I would like to thank Michele Cicoli, David Ciupke, Alexander Westphal, Luis Ib\' {a}\~{n}ez and Clemens Wieck for discussions.
This work is partially supported by the grants  FPA2015-65480-P from the MINECO/FEDER EU, the ERC Advanced Grant SPLE under contract ERC-2012-ADG-20120216-320421 and the grant SEV-2012-0249 of the ``Centro de Excelencia Severo Ochoa" Programme.

\end{document}